\documentclass[aps,prb,twocolumn,groupedaddress,draft,showpacs,intlimits,amsmath,amssymb,floats]{revtex4}

\usepackage{bm}
\usepackage[final]{graphicx}
\usepackage{epsfig}
\usepackage{subfigure}
\usepackage{verbatim}

\begin{document}


\title{Time-Dependent Charge-Order and Spin-Order Recovery in Striped Systems} 


\author{Y. F. Kung$^{1,2}$}
\author{W.-S. Lee$^{2}$}
\author{C.-C. Chen$^{3}$}
\author{A. F. Kemper$^{4}$}
\author{A. P. Sorini$^{2,5}$}
\author{B. Moritz$^{2,6}$}
\author{T. P. Devereaux$^{2,7}$}

\affiliation{$^1$Department of Physics, Stanford University, Stanford, CA 94305 USA}
\affiliation{$^2$Stanford Institute for Materials and Energy Sciences, 
SLAC National Accelerator Laboratory and Stanford University, Stanford, CA 94305, USA}
\affiliation{$^3$Advanced Photon Source, Argonne National Laboratory, Lemont, Illinois 60439, USA}
\affiliation{$^4$Lawrence Berkeley National Laboratory, 1 Cyclotron Road, Berkeley, CA 94720, USA}
\affiliation{$^5$Physics Division, Lawrence Livermore National Laboratory, Livermore, CA 94550, USA}
\affiliation{$^6$Department of Physics and Astrophysics, University of North Dakota, Grand Forks, ND 58202, USA}
\affiliation{$^7$Geballe Laboratory for Advanced Materials, Stanford University, Stanford, CA 94305, USA}

\begin{abstract}
Using time-dependent Ginzburg-Landau theory, we study the role of amplitude and phase fluctuations in the recovery of charge and spin stripe phases in response to a pump pulse that melts the orders.  For parameters relevant to the case where charge order precedes spin order thermodynamically,  amplitude recovery governs the initial time scales, while phase recovery controls behavior at longer times.  In addition to these intrinsic effects, there is a longer spin re-orientation time scale related to the scattering geometry that dominates the recovery of the spin phase.  Coupling between the charge and spin orders locks the amplitude and similarly the phase recovery, reducing the number of distinct time scales.  Our results well  reproduce the major experimental features of pump-probe x-ray diffraction measurements on the striped nickelate La$_{1.75}$Sr$_{0.25}$NiO$_4$.  They highlight the main idea of this work, which is the use of time-dependent Ginzburg-Landau theory to study systems with multiple coexisting order parameters.
\end{abstract}

\maketitle

\section{Introduction}
Strongly correlated transition-metal oxides show a variety of ordered phases, including charge density waves (CDW), spin density waves (SDW), and superconductivity, that interact to suppress or enhance one another.\cite{Dagotto_Science_2005}  For example, in the cuprate high temperature superconductors, superconductivity competes with the charge-stripe state for carriers,\cite{Tranquada_PRL_1997} as well as with the pseudogap for quasiparticles.\cite{Vishik_PNAS_2012}  In the pseudogap regime itself, resonant inelastic x-ray scattering (RIXS), angle-resolved photoemission spectroscopy (ARPES) and scanning tunneling microscopy (STM) measurements have found evidence for charge ordering.\cite{Ghiringhelli_Science_2012,Hashimoto_NatPhys_2010, Ma_PRL_2008, Vershinin_Science_2004, Wise_NatPhys_2008} Unraveling the interplay between these different orders could hold the key to understanding cuprate superconductivity.

Time-resolved pump-probe x-ray diffraction experiments at the Linac Coherent Light Source (LCLS) provide a new tool for studying how different ordered phases interact.  Excitations that are strongly coupled in equilibrium can be disentangled in the time domain, leading to new understanding of their behavior.  Because the Bragg intensity is related to both the orders' amplitudes and phases, dynamical information also allows separation of their behavior.  Disentangling these components offers insight into whether amplitude or phase fluctuations determine the strength of the order at a given time, as well as how strong coupling affects the system.

These experimental advances highlight the importance of developing a theoretical framework to analyze and interpret the experimental results.  One useful phenomenological description of these systems is the versatile time-dependent Ginzburg-Landau theory, which in principle can be applied to any coexisting or competing symmetry-broken states.  With an appropriate formulation of the order parameters, the theory can treat interactions between a wide range of phases.  Not only has it been used to model density waves, but Ginzburg-Landau theory is also used to study two-band superconductivity and $U(1)\times U(1)$ systems.\cite{Shanenko_PRL_2011,Babaev_PRL_2010}  In addition to capturing equilibrium behavior, it describes time dynamics to elucidate the effects of coupling strength between various orders, such as CDWs and superconductivity, or CDWs and SDWs in the charge-stripe state.  In order to elaborate on the specifics of these ideas, we choose to study systems with coupled charge and spin orders, which provide a connection to existing experimental efforts.

Time evolution of the order parameters calculated from the theory can be connected to experimental measurements of Bragg intensity, leading to greater understanding of how different components of phases interact.  In analogy with the atomic Bragg diffraction peaks, Bragg intensity of the ordered phases (example images shown in Fig. 1a and 1b) can be modeled in terms of each order's amplitude and electronic phase (its location with respect to the superlattice\cite{Fukuyama_PRB_1978}).  It is directly related to the squared amplitudes, but decreased by phase fluctuations, or phasons, via a Debye-Waller-like factor (Fig. 1c).\cite{Overhauser_PRB_1971}  Thus, in conjunction with experiments, time-dependent Ginzburg-Landau theory provides insights into how fluctuations of the amplitudes and phases govern the system out of equilibrium.

\begin{figure}[h!]
	\includegraphics[width=\columnwidth]{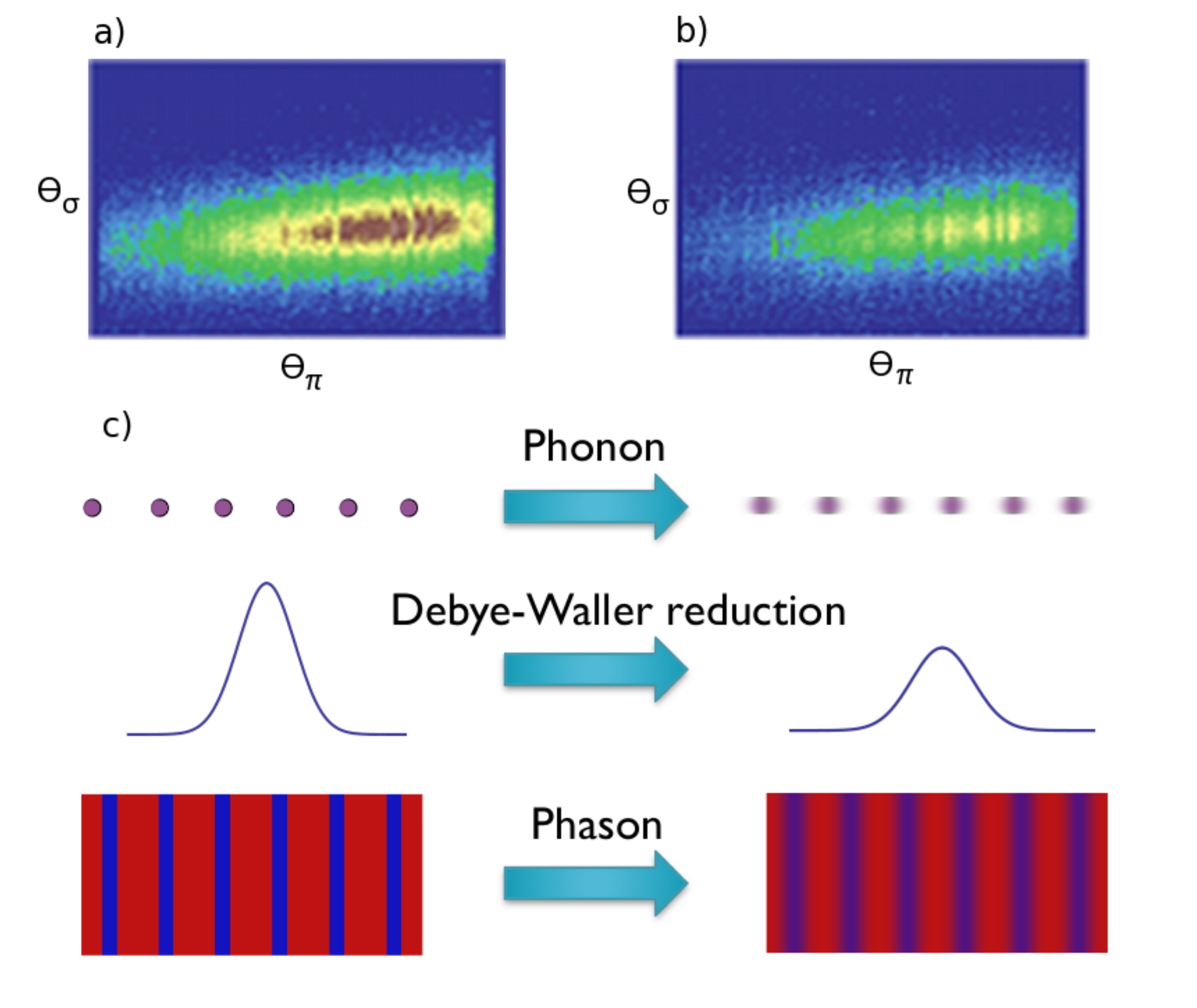}
	\caption{The experimental spin order Bragg intensity peak a) in equilibrium and b) after the pump pulse has melted the order.  The horizontal axis $\theta_\pi$ is parallel to the scattering plane, while the vertical axis $\theta_\sigma$ is perpendicular to the scattering plane.  Panel c) shows a cartoon of how phase oscillations affect Bragg intensity, in analogy with phonons.}
\end{figure}

One specific application of this theory is to CDW and SDW ordering in the nickelates,\cite{Schuessler_PRL_2005,Tranquada_Nature_1995,Cheong_PRB_1994,Tranquada_PRL_1993,Tranquada_PRL_1994} materials which are structurally similar to the cuprates and hence provide a useful model system to study interactions between these two phases.  In equilibrium, conventional CDWs may be driven by Fermi surface nesting, \cite{Gruener_RMP_1988} by strong coupling to phonons, or from an inherent instability towards phase separation driven by strong Coulomb interactions.  Most likely, the CDWs seen in the cuprates and nickelates arise due to a combination of strong coupling via both the Couloumb interaction and coupling to the lattice when holes are doped into an antiferromagnetic (AFM) parent compound: the system phase-separates into stripes of AFM domains and of greater hole density to decrease the energy penalty from breaking AFM spin interactions.\cite{Zaanen_PRB_1989, Emery_PRL_1990, Emery_PNAS_1999,Machida_JPSJ_1990,Machida_PhysicaC_1989}  This charge order is known to be strongly coupled to phonon modes.\cite{Zaanen_PRB_1994,Reznik_Nature_2006}  Strong coupling is indicated from the ratio between the CDW gap at zero temperature, $\Delta_{CDW}$, and that at $T_{CDW}$ (the critical temperature), whose measured values are 5 to 8 times larger than the weak coupling value (3.52).\cite{Varma_PRL_1983}  Since the charge-stripe state is known to interact with superconductivity in the cuprates, understanding how charge and spin orders are coupled in the nickelates may ultimately elucidate how coupling affects the critical temperature for the superconducting-normal state transition.  

In particular, the striped nickelate La$_{1.75}$Sr$_{0.25}$NiO$_4$ has demonstrated interesting cooperative dynamics between charge and spin order.\cite{Lee_NatComm_2012, Lee_PRL_2013}  After photo-excitation by a pump laser, the recovery of charge and spin order's Bragg intensities each exhibits two time scales.  Naively, one might expect charge and spin order to have different characteristic time scales, since their intrinsic energy scales differ by at least an order of magnitude.\cite{Ido_PRB_1991, Katsufuji_PRB_1996, Woo_PRB_2005}  However, the faster time scales of charge and spin order overlap.  In addition, their slow time dynamics differ by an order of magnitude, with the spin order Bragg diffraction intensity reaching a pseudo-metastable state within the experimental window.

These experimental observations highlight the need for further theoretical understanding.  In this paper, we apply the time-dependent Ginzburg-Landau theory to systems with collinear, coupled charge and spin order.  Section II explores the equilibrium features of the model, including the phase diagram.  Section III discusses its behavior in the time domain using Gross-Pitaevskii-type equations of motion.  Section IV applies the model to the case of La$_{1.75}$Sr$_{0.25}$NiO$_4$, demonstrating that it captures major features seen in the experiments and leading to the conclusion that the amplitude determines order parameter strength on fast time scales, whereas the phase controls behavior on longer time scales in charge-stripe ordered nickelates.  Finally, Section V summarizes our main findings with additional concluding remarks.

\section{Equilibrium}
The equilibrium behavior of a system with collinear stripe order can be described by the Ginzburg-Landau free energy.  We start with a free energy for a spatially varying scalar field ($\rho$) that describes charge order and a vector field ($\vec S$) that describes spin order:\cite{Zachar_PRB_1998} 
\begin{align}
F (\rho,\vec S,\theta,\phi)=\frac{1}{2}\tilde{r}_\rho |\rho|^2+|\rho|^4+\frac{1}{2}\tilde{r}_s|\vec{S}|^{2}+|\vec{S}|^4\nonumber\\
+\lambda[(\vec{S}\cdot\vec{S})\rho^{*} + h.c.]\nonumber\\
+K_\rho(\vec{\nabla} \rho^{*})\cdot(\vec{\nabla}\rho)+K_s(\vec{\nabla}|\vec{S}|^{*})\cdot(\vec{\nabla}|\vec{S}|),
\end{align}
where $\tilde{r}_\rho$ and $\tilde{r}_s$ are temperature-dependent coefficients setting the bare transition temperatures.  Close to the transition temperature, their form is given by:

\begin{align}
\tilde{r}_{\rho,s}=\alpha_{\rho,s}\frac{T-T_{\rho,s}}{T_{\rho,s}},
\end{align}
where $T_\rho$ and $T_s$ are the critical temperatures for charge and spin order, respectively, $\alpha_\rho$ and $\alpha_s$ control the frequency of vibrations if the system is perturbed, and  $\lambda[(\vec{S}\cdot\vec{S})\rho^{*} + h.c.]$ is the lowest-order term that describes coupling between spin and charge.  Higher-order terms such as $|\vec{S}|^2\rho^2$ that describe coupling between charge and spin order can be dropped as they do not lead to qualitatively different behavior.  It will be shown below that $K_\rho$ and $K_s$ are irrelevant in equilibrium.

The free energy can be simplified by re-expressing the order parameters in terms of their magnitudes and phases:

\begin{eqnarray}
\rho&=&|\rho|e^{i(\vec{Q}_\rho \cdot \vec{r}+\tilde{\theta)}}, \nonumber\\
\vec{S}&=&|S|e^{i(\vec{Q}_s \cdot \vec{r}+\tilde{\phi)}} \hat{m}.
\end{eqnarray}
Here $\vec{Q}_\rho$ and $\vec{Q}_s$ are the momentum-space ordering vectors, $\tilde{\theta}$ and $\tilde{\phi}$ are the phases, and $\hat{m}$ describes the preferred spin alignment direction.  The phases can be further expressed as a sum of modes with different momenta:

\begin{eqnarray}
\tilde{\theta}&=&\sum_q \theta_q e^{i\vec{q}\cdot\vec{r}},\nonumber\\
\tilde{\phi}&=&\sum_q \phi_q e^{i\vec{q}\cdot\vec{r}},
\end{eqnarray}
an expansion that becomes relevant when the system is excited by a pump pulse, as is the case in the time-resolved pump-probe experiments. \cite{Overhauser_PRB_1971}

The sign of the coupling term can be determined by examining the order parameters in real space:\cite{Zachar_PRB_1998}

\begin{eqnarray}
\rho(\vec{r})&\propto& |\rho|\cos{(\vec{Q}_\rho\cdot\vec{r}-\eta)},\nonumber\\
\vec{S}(\vec{r}) e^{i\vec{Q}_s\cdot\vec{r}}&\propto& |\vec{S}|\hat{m}\cos{(\vec{Q}_s\cdot\vec{r})},
\end{eqnarray}
where $\eta$ is the relative phase between the two orders.  Using these expressions and minimizing the free energy shows that if the coupling term is negative, a minimum occurs at $\eta=\pi$, whereas if the coupling term is positive, the minimum occurs at $\eta=0$.  Since doping is known to induce phase-separation into hole-rich stripes (where the CDW has its maximum) and antiferromagnetically ordered stripes (where the SDW has its maximum) in the nickelates and other correlated transition-metal oxides, the charge and spin order must have a relative $\pi$ phase shift.\cite{Zachar_PRB_1998}  Thus the coupling term is negative and stronger coupling between charge and spin order stabilizes the stripe phase.  

The gradient term simplifies to:

\begin{eqnarray}
(\vec{\nabla} \rho^{*})\cdot(\vec{\nabla}\rho)&=&(\vec{\nabla}|\rho|)^2+|\rho|^2|\vec{Q}_\rho|^2\nonumber\\
&&+|\rho|^2\sum_q |\vec{q}|^2|\theta_q|^2,
\end{eqnarray}
and similarly for spin.  Assuming that the magnitudes of the order parameters are spatially invariant for systems of interest, such as the nickelates,\cite{Lee_NatComm_2012} the $(\vec{\nabla}|\rho|)^2$ term vanishes, and the second term can be combined with the $\tilde{r}_\rho$ term, such that $r_\rho=\tilde{r}_\rho+|\vec{Q}_\rho|^2$.  In equilibrium, intuition dictates that the real-space phases are fixed and defined to be 0, making the phase diagram invariant with respect to the final term.  These terms for both the charge and spin orders would in principle impose an energy penalty on phase fluctuations, which are neglected in equilibrium.

Finally the equilibrium free energy for the striped system can be written as

\begin{figure}[h!]
	\includegraphics[width=\columnwidth]{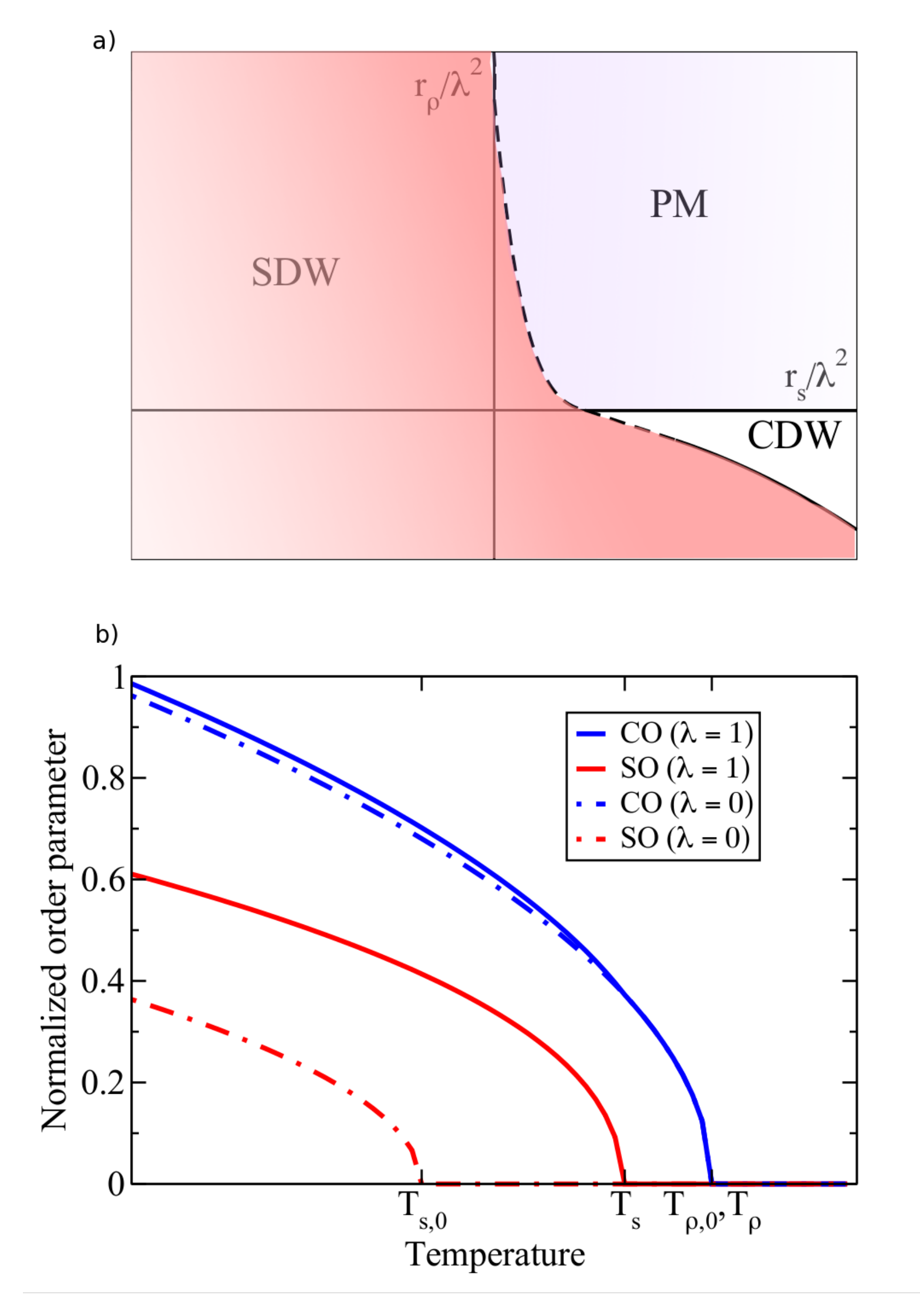}
	\caption{a) Equilibrium phase diagram determined by $r_\rho/\lambda^2$ and $r_s/\lambda^2$, with the first order transitions indicated by a dashed line and second order transitions marked by solid lines.  b) The phase diagram of charge order (CO) and spin order (SO) versus temperature, demonstrating second-order phase transitions and the stabilization of spin order via coupling to charge order, for representative parameters.  The uncoupled spin order critical temperature $T_{s,0}$ is increased significantly when the order parameters are coupled.  However, $T_{\rho,0}$ does not change noticeably, indicating that charge order is the more thermodynamically stable state.  All order parameter values are normalized to the value of CO at $T=0$, with coupling.}
\end{figure}
\begin{eqnarray}
F (|\rho|,|S|,\theta,\phi)&=&\frac{1}{2}r_\rho |\rho|^2 +|\rho|^4+\frac{1}{2}r_s|S|^{2}+|S|^4\nonumber\\
&&-2\lambda|S|^2|\rho|,
\end{eqnarray}
In addition, it can be seen that the coupling term is only allowed by symmetry when $\vec{Q}_\rho=2\vec{Q}_s$, enforcing a $2:1$ relationship between the charge and spin momentum space vectors that is indeed seen in materials such as the nickelates and cuprates.  

Fig. 2a shows a generic phase diagram for this free energy, illustrating how $r_\rho$ and $r_s$, normalized by $\lambda^2$, determine the phase of the system.  There is a first order transition (dashed line) between  the paramagnetic (PM) phase and SDW phase, which extends over the second and third quadrants, as well as parts of the first and fourth.  The CDW phase exists in the fourth quadrant and is separated from the SDW phase by either a first order transition (dashed line) or second order transition (solid line).  The SDW phase exhibits charge ordering in addition to spin ordering.\cite{Zachar_PRB_1998,Balents_PRL_2011}

Focusing on regions of the phase diagram where the CDW-SDW transition is second order, Fig. 2b shows the temperature dependence of the charge and spin order parameters for representative values of $r_\rho$ and $r_s$.  Coupling the orders significantly increases the spin order critical temperature from $T_{s,0}$ to $T_s$, as well as the value of the order parameter itself, indicating that it becomes more stable.  Meanwhile, neither the critical temperature nor the order parameter of charge order changes noticeably when the orders are coupled.  This behavior is consistent with charge order being more thermodynamically stable, which is known to be the case in the nickelates.\cite{Yoshizawa_PRB_2000}

\section{Time evolution}

The time evolution of the charge and spin order parameters can be determined from Gross-Pitaevskii-type equations of motion\cite{Gross_NuovoCimento_1961} derived from the Lagrangian of the system.  The order parameters' kinetic energy is written as $\dot{\rho}^2+\dot{S}^2=|\dot{\rho}|^2+|\dot{S}|^2+|\rho|^2\dot{\theta}^2+|S|^2\dot{\phi}^2$, and the Lagrangian is found by subtracting the free energy $F$ from the kinetic energy.  When the system is driven out of equilibrium, the phase fluctuations play a role in the dynamics, so the free energy in Eq. 7 must be modified.  Depending on its energy, the pump excites certain phase modes.  For example, if an optical pump laser is used, only low-momentum phase modes are excited, and their behavior can be described by an effective phason for charge order ($\theta$) and spin order ($\phi$).  The coupling term then becomes $-2\lambda|S|^2|\rho|\cos{(2\phi-\theta)}$.  In addition, the final term from the gradient (see Eq. 6) is included as $K_\theta |\rho|^2\theta^2+K_\phi |S|^2\phi^2$, where the factor of $|\vec{q}|^2$ has been absorbed into the coefficients.  (In cases where the amplitudes are perturbed only slightly from equilibrium, these terms can be further approximated as $K^{'}_\theta \theta^2+K^{'}_\phi \phi^2$.)

To account for energy loss, the Euler-Lagrange equations of motion are supplemented with phenomenological damping terms proportional to the first derivatives of the order parameters.  Incorporating these effects, the behavior of the system is described by the equations of motion:

\begin{eqnarray}
\ddot{|\rho|}& = &|\rho|\dot{\theta}^2-\frac{1}{2}r_\rho|\rho|-2|\rho|^3\nonumber\\
&&+\lambda|S|^2\cos{(2\phi-\theta)}-2K_\theta |\rho|\theta^2-\gamma_\rho\dot{|\rho|},\nonumber\\
\ddot{|S|}& = &|S|\dot{\phi}^2-\frac{1}{2}r_s|S|-2|S|^3\nonumber\\
&&+2\lambda|S||\rho|\cos{(2\phi-\theta)}-2K_\phi |S|\phi^2-\gamma_S\dot{|S|},\nonumber\\
\ddot{\theta} &=& 2\lambda|S|^2|\rho|\sin{(2\phi-\theta)}-2K_\theta |\rho|^2\theta-\gamma_\theta\dot{\theta},\nonumber\\
\ddot{\phi}& =& -4\lambda|S|^2|\rho|\sin{(2\phi-\theta)}-2K_\phi |S|^2\phi-\gamma_\phi\dot{\phi}.
\end{eqnarray}
Solving these four coupled differential equations determines the time dependence of the order parameters.

Examining the amplitude fluctuations alone elucidates their effect on the stability of the order.  The dynamics of the amplitude --- via the amplitudons --- are governed by the equations:

\begin{eqnarray}
\ddot{|\rho|}&=&-\frac{1}{2}r_\rho|\rho|-2|\rho|^3+\lambda|S|^2 -\gamma_\rho\dot{|\rho|},\nonumber\\
\ddot{|S|}&=&-\frac{1}{2}r_s|S|-2|S|^3+2\lambda|S||\rho|-\gamma_S\dot{|S|}.
\end{eqnarray}
When the uncoupled and undamped order parameters are perturbed slightly, $|\rho|$ and $|S|$ will oscillate with frequencies proportional to $\sqrt{|r_\rho|}$ for charge order and $\sqrt{|r_s|}$ for spin order.  Coupling the two orders will increase the amplitude oscillation frequency of both, although similar to the equilibrium case, charge order is more robust than spin order against amplitude fluctuations.

If the regime of interest involves longer time scales, with the return to equilibrium dominated by lattice coupling or spin-orbit effects, the second order derivatives in the equations of motion can be dropped to focus on decay at longer times.  In this regime, small perturbations from equilibrium decay according to the equations of motion:

\begin{eqnarray}
0&=&-\frac{1}{2}r_\rho|\rho|+\lambda|S|^2 -\gamma_\rho\dot{|\rho|},\nonumber\\
0&=&-\frac{1}{2}r_s|S|+2\lambda|S||\rho|-\gamma_S\dot{|S|}.
\end{eqnarray}
In the uncoupled case, $\lambda=0$ and the exact decay time scales are found (analytically) to be $\tau_\rho=\gamma_\rho/|r_\rho|$ and $\tau_s=\gamma_s/|r_s|$ for charge and spin order, respectively.

When the assumption of small perturbations is relaxed, the cubic terms are restored to the equations of motion, but the uncoupled case can still be solved analytically.  If the pump pulse initially suppresses the charge order magnitude by an amount $\Delta\rho$ (not necessarily small), the order recovers with a rate of 

\begin{align}
\tau_\rho=\frac{2\gamma_\rho}{|r_\rho|}\left(\ln\left[\frac{1-\Delta\rho/e}{1-\Delta\rho}\right]-\frac{1}{2}\ln\left[\frac{1-(1-\Delta\rho/e)^2}{1-(1-\Delta\rho)^2}\right]\right),
\end{align}
where $e$ is the Euler constant.  Similarly for spin order, when the magnitude is suppressed by $\Delta S$,

\begin{align}
\tau_s=\frac{2\gamma_s}{|r_s|}\left(\ln\left[\frac{1-\Delta S/e}{1-\Delta S}\right]-\frac{1}{2}\ln\left[\frac{1-(1-\Delta S/e)^2}{1-(1-\Delta S)^2}\right]\right).
\end{align}
The amplitude recovery time scales are proportional to the damping strength $\gamma_{\rho,s}$ and inversely proportional to the stability of the order $|r_{\rho,s}|$.  This shows that the more robust the order, the faster the amplitudons decay, so that the system returns more quickly to equilibrium.

The role of phasons in the recovery process can also be studied using Eq. (8), where the second derivative terms are again dropped to focus on recovery at longer times.   As discussed in Section II, the phase that enters into the Ginzburg-Landau theory is an effective quantity that encompasses the low energy phase modes excited by the optical pump laser.  The phason dispersion is gapless whereas that of amplitudons is gapped, so the low energy phase modes considered here always have a lower energy\cite{Ren_JChemPhys_2004} and hence govern behavior on longer time scales than the amplitudons.  Phason recovery also will be slower than spin order amplitude recovery, since coupling of spin order to the more robust charge order decreases the spin order amplitude time scale.

%
%

To simulate time-resolved x-ray diffraction experiments that probe the behavior of systems with collinear charge and spin order, the amplitude and phase order parameters must be related to the Bragg intensity.  In analogy with the Bragg peaks from a crystalline lattice, the Bragg peaks of the charge and spin order superlattices are proportional to the squared amplitude and a Debye-Waller-like factor from the effective phase.\cite{Overhauser_PRB_1971}  Just as incoherent motion of atoms about their equilibrium lattice positions leads to a reduction of the measured intensity, so fluctuations of the phase decrease the charge and spin order Bragg intensity.  Incorporating this effect, the normalized Bragg intensities can be written in terms of the order parameters as:

\begin{eqnarray}
I_{CO}&=&\frac{|\rho|^2}{\rho_0^2}e^{-\theta^2},\nonumber\\
I_{SO}&=&\frac{|S|^2}{S_0^2}e^{-\phi^2},
\end{eqnarray}
where $\rho_0$ and $S_0$ are the equilibrium values.

A final effect on the measured Bragg intensity of spin order only can come from the experimental scattering geometry.  Due to spin order's vector nature, its intensity can be modified by the factor:\cite{Hannon_PRL_1988}

\begin{align}
G=|(\epsilon_{in}\times\epsilon_{out})\cdot \hat{m}|^2,
\end{align}
where $\epsilon_{in}$ gives the polarization of the incoming probe pulses, $\epsilon_{out}$ gives the polarization of the diffracted light, and $\hat{m}$ is the direction of spin order.  In pump-probe experiments, the pump pulse perturbs the system and rotates the spins away from their equilibrium orientation, changing $G$ and affecting the measured intensity.  Since this effect occurs for vector order parameters only, charge order is unaffected, but the expression for spin order Bragg intensity should be modified to:

\begin{align}
I_{SO}=\frac{|S|^2}{S_0^2}e^{-\phi^2}G
\end{align}
where relevant.  The time scale for $G$ is related to the spin order vector re-orientation and hence to spin-orbit coupling.

As an interesting note, the pump pulse can in principle rotate the spins to a new orientation such that $G$ is either less or greater than it was in equilibrium.  In the former case, the spin order Bragg intensity will be suppressed, and it will attain a metastable value less than the equilibrium value.  On the other hand, in the latter case, an enhancement of spin order Bragg intensity should be observed, and it will recover to a metastable value \emph{greater} than its equilibrium value.  However, experiments typically optimize the geometry to maximize the equilibrium Bragg intensity, so it is unlikely to see the enhanced metastable state.

\section{Application to Striped Nickelate}
The time-dependent Ginzburg-Landau model of a charge and spin striped system can be applied to the nickelate La$_{1.75}$Sr$_{0.25}$NiO$_4$, with experimental considerations guiding the choice of parameters.  Optical conductivity experiments have measured a charge transfer energy on the order of 1 eV and a charge gap of 200 meV,\cite{Ido_PRB_1991,Katsufuji_PRB_1996} whereas inelastic neutron scattering experiments have found a spin superexchange energy of 10-30 meV and a spin gap of 20 meV. \cite{Woo_PRB_2005}  Together, these results suggest that charge order has an intrinsic energy scale that is at least an order of magnitude greater than that of spin.  

Using these considerations, we set $r_\rho\sim10 r_s$ ($r_\rho=-105$ and $r_s=-10$).  These values guarantee that the phase transitions are second order, as is known in the nickelates.\cite{Tranquada_Nature_1995}  In addition, it elucidates the behavior of the system when charge order's energy scale is an order of magnitude stronger than spin's in the absence of coupling.  As discussed already, the phase fluctuations have low energy compared to the charge order amplitudons, motivating the choice of $K_\theta=K_\phi=0.15\sim |r_s|/100$. 

Since strong electron-lattice effects are important to forming the striped state in the nickelates, \cite{McQueeney_PRL_1999, McQueeney_PRB_1999, Hotta_PRL_2004} charge and spin order dissipate energy via the lattice, as captured by the damping terms in the equations of motion.  Charge order couples directly to the lattice though the Coulomb interaction, whereas spins couple only indirectly,\cite{Ament_RMP_2011} so charge order has a stronger relaxation pathway and $\gamma_\rho>\gamma_s$.   The low-energy phase fluctuations have a correspondingly long time scale, so phase recovery is chosen to take at least an order of magnitude longer than amplitude recovery in the uncoupled case.  The damping factors are selected such that $\gamma_\theta=\gamma_\phi\sim 10\gamma_\rho$ ($\gamma_\rho=150$, $\gamma_s=50$, $\gamma_\theta=\gamma_\phi=1000$).  With these parameters, the uncoupled amplitude decay time scale for charge order is a few times that of spin order, highlighting the effect of coupling on the time scales. 

The model can be tested for different strengths of the pump pulse that melts the order.  For weak pump fluence, the initial conditions are perturbed slightly from equilibrium to be $|\rho(0)|=0.8\rho_0$ and $|S(0)|=0.9S_0$, where $\rho_0$ and $S_0$ are the equilibrium values.  The ratio of the phases is assumed to remain close to the $2:1$ equilibrium ratio ($\theta(0)=0.45$ rad and $\phi(0)=0.2$ rad).  Stronger pump fluence should disrupt the stripe phase to a greater degree, so the initial amplitudes are chosen to be suppressed more ($|\rho(0)|=0.6\rho_0$ and $|S(0)|=0.7S_0$).  The ratio of phases is chosen as $\theta(0)=0.9$ rad and $\phi(0)=0.2$ rad, differing from the equilibrium ratio by a factor of two to test the effect of significant phase disorganization.  Finally, in order to simulate the temporal experimental resolution, the calculated Bragg intensity time traces are convolved with a Gaussian with a standard deviation of 1.69 arbitrary time units (value chosen to agree qualitatively with experiment).

\begin{figure}[h!]
\includegraphics[width=\columnwidth]{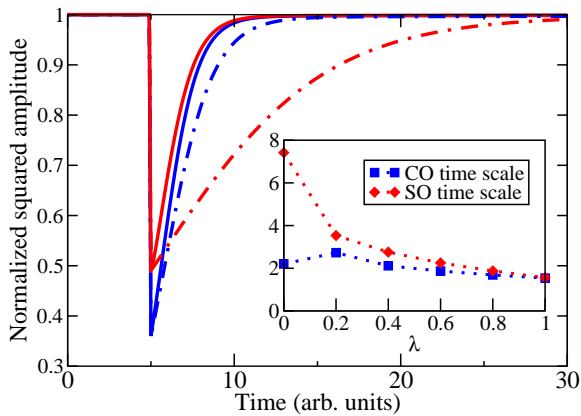}
	\caption{The normalized squared amplitude time traces for charge order in the coupled (solid blue line) and uncoupled (dashed blue line) cases, and for spin order in the coupled (solid red line) and uncoupled (dashed red line) cases.  The inset shows the initial recovery time scales (in arbitrary units) for different coupling strengths, with $\lambda$ normalized such that $\lambda=1$ is strong coupling.}
\end{figure}

The time scales associated with amplitude recovery can be extracted by fitting the calculated Bragg intensity curves with an exponential.  As discussed already, a priori the non-interacting time scales are expected to be intrinsically different.  In the uncoupled model ($\lambda=0$), unless the parameters are very carefully tuned, the recovery time scales are different, as illustrated in Fig. 3 (inset).  However, as soon as the order parameters are coupled, any nonzero value of the coupling constant $\lambda$ will begin to bring the time scales together.  As the coupling strength is increased, the time scales approach each other more closely until the two orders lock to each other.  Importantly, once locked, they do not separate again with a further increase in coupling.  Fig. 3 also shows that the amplitudes rapidly recover to their equilibrium values within the time window studied, implying that the phases govern the longer time recovery, which is expected from physical arguments as discussed above.  

This phase relaxation is shown in the inset of Fig. 4 for the strong pump fluence case (the weak pump fluence case behaves similarly).  Due to the coupling term in the free energy that is proportional to $\cos{(2\phi - \theta)}$, finite coupling quickly locks the phases together such that $2\phi-\theta=2\pi n$, with integer $n$.  The $n=0$ case is most likely, since $n>1$ requires larger effective phases and hence higher energy phason modes.  This behavior can be seen in the time evolution of the effective charge and spin orders: although the ratio of initial phases is very different from its equilibrium ratio, coupling quickly locks the charge and spin orders into a $2:1$ relationship, after which they slowly relax back to equilibrium together.  
\begin{figure}[h!]
	\includegraphics[width=\columnwidth]{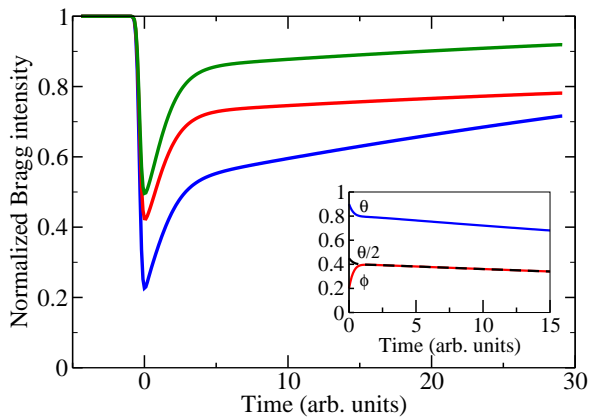}
	\caption{The calculated Bragg intensity time traces for charge order (blue), vector spin order (red), and scalar spin order (green), illustrating the effect of the geometric scattering factor on spin metastability.  The inset shows that the phases of charge and spin order lock together rapidly in a $2:1$ ratio before relaxing slowly to equilibrium together, demonstrating the need for physics beyond phase evolution to describe the spin metastability.}
\end{figure}

However, since this locking process occurs on a shorter time scale than that of the spin metastability, phase recovery alone cannot account for the suppression of the Bragg intensity observed in experiments.  Hence spin order's orientation must affect the measured intensity on long time scales.  The spin-orbit coupling in the nickelates is weak,\cite{Kuiper_PRB_1998} so its effect is to pick an easy axis, indicating the smallness of the associated energy scale and hence the slowness of the corresponding time scale, which is the longest in the system.  This effect from spin order's vectorial nature strongly affects recovery of the Bragg intensity via the experimental scattering geometry factor $G$.  Because the spin order re-orientation is the longest time scale in the system, it can be assumed to remain constant during times of interest.  A weak pump pulse alters spin order direction less than a strong pump pulse, so we take $G=0.94$ in the former case and $G=0.85$ in the latter.  The calculated Bragg intensities of charge and spin order for vector and scalar spin order demonstrate that spin re-orientation is what gives rise to the spin metastability (see Fig. 4).  (The model shows qualitatively similar behavior for other values of the geometrical scattering factor; these specific values were chosen to give qualitatively similar results as the experiment.)  We stress that the inclusion of this term has no effect on the amplitude and phase dynamics; its only effect is on the Bragg intensity.

\begin{figure}[h!]
	\includegraphics[width=\columnwidth]{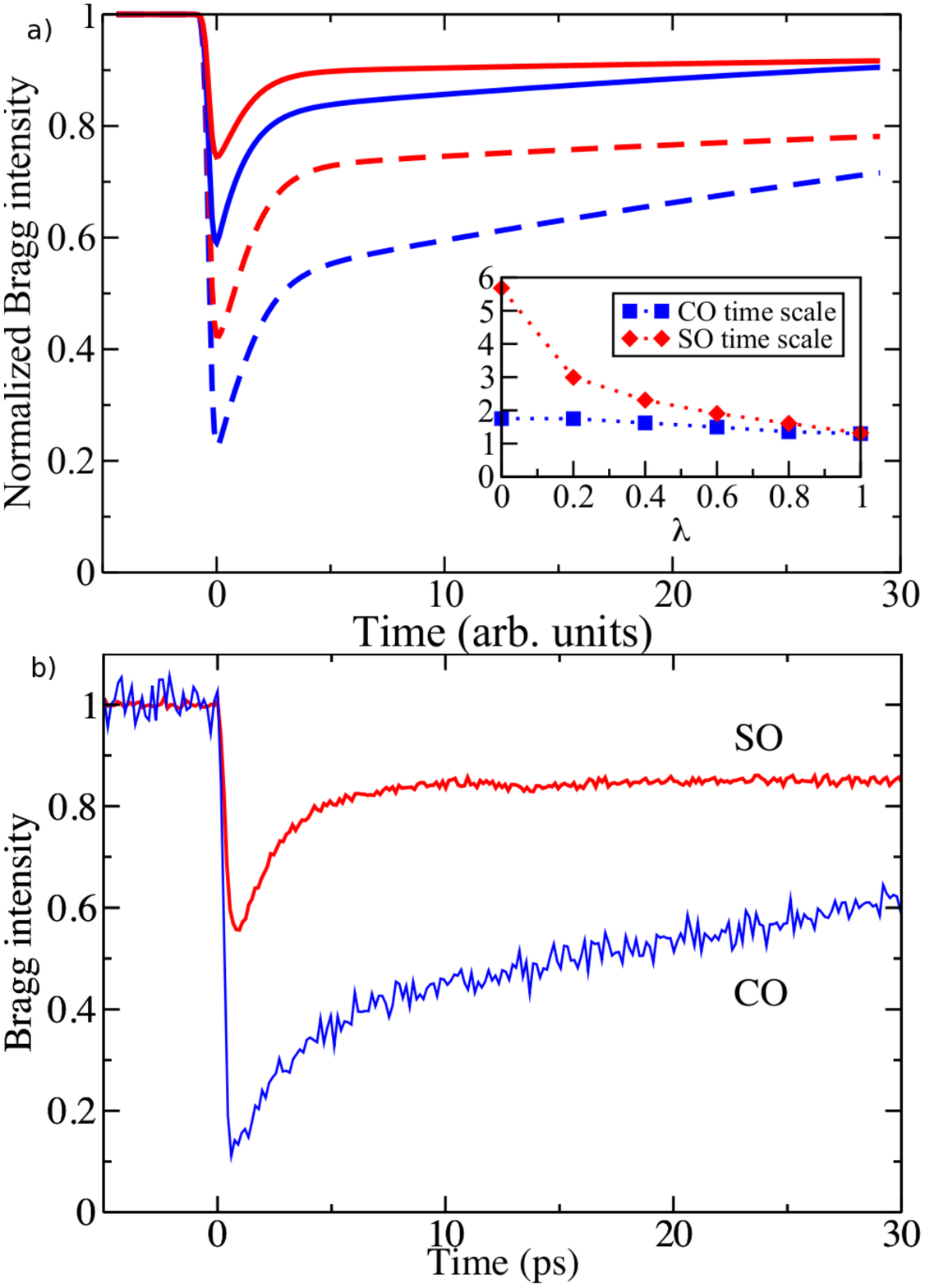}
	\caption{Bragg intensity time traces for charge (blue) and spin (red) order calculated from theory for two different pump strengths a) compared to experiment b).  The experimental data for spin order were taken with an excitation density of 38 J/cm$^2$, whereas the excitation density for the charge order measurements was 29 J/cm$^2$.  The inset in a) shows the locking of the initial recovery time scales in the weak pump pulse case as coupling between the orders increases.}
\end{figure}

Finally, all these considerations can be incorporated to show simulated Bragg intensity time evolution in the cases of weak and strong pump fluence.  Figure 5 demonstrates the qualitative similarity of the simulated time traces (Fig. 5a) for both sets of initial conditions (when coupling between charge order and spin order is finite) to the experimental time traces (Fig. 5b).  As discussed already, coupling between charge and spin order locks their faster, initial recovery dynamics together.  This overlap can be seen from Fig. 5a, but a two-time fit is also performed on the simulated Bragg intensity time traces to extract the time scales.  For the weak pump pulse case, these are plotted in the inset to show the effect of increasing coupling strength.  On longer times, the recovery of both orders slows down as the phase recovers, and spin order reaches a metastable state due to the long time scale of spin re-orientation.

At a glance, there appears to be five distinct time scales in the system: two for the order parameters' amplitudes, two for the effective phases, and one for the vector re-orientation of spin order.  However, coupling between charge and spin order locks together the amplitude dynamics, and also the phase dynamics, reducing the number of time scales to the three seen in experiment.  Thus, the time-dependent Ginzburg-Landau model describes the full temporal dynamics of the pump-probe experiment and elucidates the role of amplitude and phase fluctuations, as well as of spin re-orientation, in the recovery process.

As a note, the qualitative behavior of the model is relatively insensitive to parameter choice (as long as they satisfy the conditions of thermal equilibrium).  For example, decreasing $r_\rho$ and $r_s$ each by a factor of 10 will change the initial recovery time scales for both the charge and spin order Bragg intensities, but they remain locked to each other as long as coupling strength is finite.  Changing the ratio of amplitude damping factors by 10 and increasing the coupling strength leads to similar conclusions.  In addition, varying the phase damping factors alters the rate at which the Bragg intensities recover at longer times, but does not affect the locking of the initial recovery.  This robustness demonstrates that the observation of finite coupling driving the amplitude recovery time scales together is not specific to a particular parameter set, but rather is a general property of this model.

\section{Conclusions}
The time-dependent Ginzburg-Landau model qualitatively captures the time evolution of coupled, collinear charge and spin order parameters in response to a pump pulse that melts the equilibrium stripe order.  With the appropriate choice of parameters, it shows that on short time scales, the stability of the order parameters is determined by the amplitude fluctuations, whereas on longer time scales, phase fluctuations dominate.  Interestingly, despite being melted by the pump pulse, the charge and spin order parameter dynamics do not decouple even out of equilibrium: instead, they are governed by the joined time scales determined by the choice of parameters.  

This result is surprising, given the significant difference in characteristic energy scales and hence the intuitively expected time scales between charge and spin.  It can be reconciled by studying the effect of coupling between the order parameters, which exists even out of equilibrium and is what locks the recovery dynamics together.  Because coupling acts on a macroscopic scale to intertwine the time scales, it does not affect the intrinsic microscopic energy scales.  

Additionally, the model incorporates the effect of scattering geometry on the measured Bragg diffraction intensity, showing that on long time scales, spin order recovery is dominated by a slow spin re-orientation process that can mask the spin phase time scale and lead to the observed spin metastability.  Thus finite coupling between charge and spin order explains the reduction of five distinct expected time scales (two for the amplitude recovery, two for phase recovery, and one for spin re-orientation) to the three observed in experiment.  

The time-resolved pump-probe experiments have demonstrated the need for a new way of thinking about collective many-body dynamics.  In systems such as the strongly correlated transition-metal oxides, studying the charge, spin, orbital, and lattice degrees of freedom separately is insufficient to capture their behavior.  Instead, strong interactions between the different degrees of freedom necessitate treating them on equal footing in the theoretical framework, and studying the dynamics of coupled orders rather than individual ones.  Though we have applied it specifically to La$_{1.75}$Sr$_{0.25}$NiO$_4$, the phenomenological Ginzburg-Landau model provides a general way of understanding the effect of strong coupling in striped, collinear systems.  In fact, with the appropriate definition of order parameters, the approach can be generalized to describe the time dynamics of systems with any coupled, symmetry-broken states, such as density waves, two-band superconductivity, and two-component superfluidity.

We would like to thank S. A. Kivelson and M. W. Haverkort for helpful discussions.  This research was supported by the U.S. Department of Energy, Office of Basic Energy Sciences, Division of Materials Sciences and Engineering, under Contract No. DE-AC02-76SF00515, SLAC National Accelerator Laboratory (SLAC), Stanford Institute for Materials and Energy Science.  Y. F. K. is supported by the Department of Defense (DoD) through the National Defense Science and Engineering Graduate (NDSEG) Fellowship.  C.C.C. is supported by the Aneesur Rahman Postdoctoral Fellowship at Argonne National Laboratory, operated under the U.S. DOE contract No. DE-AC02-06CH11357.

\bibliography{Nickelate_Bib}

\end{document}